\pdfoutput=1
\documentclass[sigconf]{acmart}

\usepackage{booktabs} 
\usepackage{amsmath}
\usepackage[printonlyused]{acronym}
\usepackage{paralist}
\usepackage[shortlabels]{enumitem}
\usepackage[skip=0pt]{caption}
\usepackage{subcaption}

\copyrightyear{2019}
\acmYear{2019} 
\setcopyright{iw3c2w3}
\acmConference[WWW '19]{Proceedings of the 2019 World Wide Web Conference}{May 13--17, 2019}{San Francisco, CA, USA}
\acmBooktitle{Proceedings of the 2019 World Wide Web Conference (WWW'19), May 13--17, 2019, San Francisco, CA, USA}
\acmPrice{}
\acmDOI{10.1145/3308558.3313415}
\acmISBN{978-1-4503-6674-8/19/05}

\usepackage{balance}

\newcommand{\CPF}{CP$_{\mathit{free}}$}

\DeclareMathOperator*{\argmax}{arg\,max}

\DeclareMathOperator*{\freq}{freq}

\DeclareMathOperator*{\TL}{TL}
\DeclareMathOperator*{\WTL}{WTL}
\DeclareMathOperator*{\CE}{CE}
\DeclareMathOperator*{\CP}{CP}

\DeclareMathOperator*{\FACE}{FACE}

\acrodef{Seq2Seq}{Sequence-to-Sequence}
\acrodef{WLF}{Weighted Loss Function}
\acrodef{NN}{Neural Network}
\acrodef{RNN}{Recurrent Neural Network}
\acrodef{FACE}{Frequency-Aware Cross-Entropy}
\acrodef{CE}{Cross-Entropy}
\acrodef{MMI}{Maximum Mutual Information}
\acrodef{MHAM}{Multi-Head Attention Mechanism}
\acrodef{VAE}{Variational AutoEncoder}
\acrodef{MAP}{max a posteriori}
\acrodef{LSTM}{Long Short-Term Memory}
\acrodef{TL}{total loss}
\acrodef{MLP}{Multi-Layer Perceptron}
\acrodef{CP}{Confidence Penalty}
\acrodef{HS}{hidden size}
\acrodef{RF}{relative frequency}
\acrodef{OSDb}{OpenSubtitles database}

\begin{document}

\title[Frequency-Aware Cross-Entropy Loss]{Improving Neural Response Diversity with Frequency-Aware Cross-Entropy Loss}

\author{Shaojie Jiang}
\affiliation{%
  \institution{University of Amsterdam}
  \city{Amsterdam}
  \state{The Netherlands}
}
\email{s.jiang@uva.nl}

\author{Pengjie Ren}
\affiliation{%
  \institution{University of Amsterdam}
  \city{Amsterdam}
  \state{The Netherlands}
}
\email{p.ren@uva.nl}

\author{Christof Monz}
\affiliation{%
  \institution{University of Amsterdam}
  \city{Amsterdam}
  \state{The Netherlands}
}
\email{c.monz@uva.nl}

\author{Maarten de Rijke}
\affiliation{%
  \institution{University of Amsterdam}
  \city{Amsterdam}
  \state{The Netherlands}
}
\email{derijke@uva.nl}

\begin{abstract}
\ac{Seq2Seq} models have achieved encouraging performance on the dialogue response generation task.
However, existing \ac{Seq2Seq}-based response generation methods suffer from a \emph{low-diversity problem}: they frequently generate generic responses, which make the conversation less interesting.
In this paper, we address the low-diversity problem by investigating its connection with \emph{model over-confidence} reflected in predicted distributions.
Specifically, we first analyze the influence of the commonly used \ac{CE} loss function, and find that the \ac{CE} loss function prefers high-frequency tokens, which results in low-diversity responses.
We then propose a \ac{FACE} loss function that improves over the \ac{CE} loss function by incorporating a weighting mechanism conditioned on token frequency.
Extensive experiments on benchmark datasets show that the \ac{FACE} loss function is able to substantially improve the diversity of existing state-of-the-art \ac{Seq2Seq} response generation methods, in terms of both automatic and human evaluations.
\end{abstract}

%
%



\maketitle


\section{Introduction}

Recently, dialogue response generation has attracted a lot of attention due to its potential for applications, e.g., within intelligent customer service agents and personal assistants.
Most state-of-the-art approaches to this task are based on \acfi{Seq2Seq} frameworks \citep{vinyals2015neural,li2016diversity,sordoni2015neural,serban2016building,serban2017hierarchical,zhao2017learning}.
Current \ac{Seq2Seq} frameworks suffer from a \emph{low-diversity problem}.
As a result, these approaches frequently generate generic responses such as ``\emph{I don't know}'' or ``\emph{I'm sorry}'' \citep{li2016diversity,serban2016building,serban2017hierarchical,zhao2017learning}.

To address the issue of low response diversity, previous studies have produced several hypotheses about the cause of the low-diversity problem with corresponding solutions.
For example, \citet{li2016diversity} argue that the usual \ac{MAP} objective function might favor frequent responses.
Instead, they propose to use a \ac{MMI} objective function to encourage responses with higher mutual information regarding the user's utterance.
The downside of this method is that it relies heavily on beam search and/or an inverse model that is trained by swapping inputs and outputs.
\citet{tao2018get} assume that the single attention layer commonly used in \ac{Seq2Seq} models is only able to focus on a single semantic aspect of the input sequence, so they propose a \ac{MHAM} to allow the decoder to generate more diverse responses.
There are also studies that try to improve response diversity by introducing randomness \citep{serban2017hierarchical,zhao2017learning}.

Recently, \citet{jiang2018why} have shown that there is a strong connection between the low-diversity problem and what they call \emph{model over-confidence} following~\citet{pereyra2017regularizing}, the phenomenon that a model incorrectly assigns most probability to only a few tokens.
However, the cause of this phenomenon remains unknown.
In \S\ref{analysis}, we investigate and conclude that the model over-confidence problem is caused by an imbalanced training of tokens with variable frequencies, which favors frequent tokens and results in low-diversity.
To correct for this, we propose a \acfi{FACE} loss function that improves the traditional \acfi{CE} loss function by taking token frequency into consideration.
More specifically, we first analyze the influence of the commonly used \ac{CE} loss function, and find that it prefers high-frequency tokens, which results in model over-confidence and low-diversity responses.
Then we propose a \ac{FACE} loss function that improves over the \ac{CE} loss function by incorporating a weighting mechanism conditioned on token frequency.

The primary differences between \ac{FACE} and previous studies in addressing the low-diversity problem are two-fold:
\begin{inparaenum}[(1)]
\item The diversity improvements brought by \ac{FACE} do not rely on beam search or randomness; and
\item \ac{FACE} does not introduce new layers or hyper-parameters to \ac{Seq2Seq} models.
\end{inparaenum}

We perform extensive experiments on two benchmark datasets, namely \ac{OSDb} and Twitter.
Compared with deterministic \ac{Seq2Seq}-based methods, like \ac{MMI} \citep{li2016diversity} and \ac{MHAM} \citep{tao2018get}, \ac{FACE} achieves the highest diversity performance and it does so with minimum modifications to the original \ac{Seq2Seq} model structure and existing hyper-parameters.\footnote{Source code for both \ac{FACE} and the baselines, together with the validation and test sets used in our experiments can be found at https://github.com/ShaojieJiang/FACE.}

The main contributions we make in this paper are:
\begin{itemize}[leftmargin=*,nosep]
\item We examine the influence of token frequency on model over-confidence and response diversity.
\item We propose a \acfi{FACE} loss function to balance the per-token loss, which alleviates model over-confidence and, hence, improves response diversity.
\item We investigate two token frequency calculation methods and corresponding frequency-based weighting mechanisms for \ac{FACE}.
\end{itemize}



\section{Low Diversity, Model Over-confidence and Loss Imbalance}
\label{analysis}

As illustrated by \citet{jiang2018why}, model over-confidence \citep{pereyra2017regularizing} can cause \ac{Seq2Seq}-based conversation models to have low response diversity.
In this section, we show that model over-confidence and low-diversity are statistical and empirical symptoms of the same problem. 
Imbalanced training is the actual underlying reason.

\begin{table}
  \centering
  \caption{Frequency ranks of leading tokens and their percentage (\%). Validation rank and Test rank are for the validation and test model responses, Training rank is for the ground-truth responses in the training data.}
  \label{tab:freq}
  \begin{tabular}{lcccccc}
    \toprule
    & i & the & you & we & he & it \\
    \midrule
    Validation rank & 1 (74) & \phantom{1}2 (5) & 3 (5) & 4 (3) & 5 (3) & 6 (3)\\
    Test rank & 1 (72) & \phantom{1}2 (7) & 3 (5) & 5 (3) & 6 (3) & 4 (2) \\
    Training rank & 1 (14) & 7 (12) & 3 (6) & 6 (3) & 8 (3) & 4 (3) \\
    \bottomrule
  \end{tabular}
\end{table}

Existing \ac{Seq2Seq}-based conversation models tend to be over-confident during prediction and place most probability on only a few tokens.
We define a \emph{leading} token in a response to be a token that appears first.
From Table~\ref{tab:freq}, we can see that all of the most frequent leading tokens in model responses are actually very likely to appear first in training ground-truth responses, but the percentages of the top-ranked token ``i'' are much higher in the validation and test sets.
This suggests that the model is over-fitting for frequent leading tokens.
We also find that the frequency of some subsequent tokens (e.g., \emph{'m} and \emph{not}) is much higher in model responses than in the training ground truth, as illustrated in Table~\ref{tab:lm}.
This is also due to model over-fitting, since these tokens are \emph{relatively} more frequent compared to other tokens.
For example, given that the previous token is ``i'', the frequency of the following token ``'m'' is much higher than others in the training data.

To understand this phenomenon and introduce our solution, we first look into the commonly used \acfi{MAP} objective function of \ac{Seq2Seq} models.
Given a dataset of message-response pairs $(X, Y)$, where $X=(x_1, x_2, \dots, x_{|X|})$ and $Y=(y_1, y_2, \dots, y_{|Y|})$ are the input and output sequences, respectively, the goal of \ac{Seq2Seq} training is to maximize the conditional probability $P(Y|X)$.
Since the decoder \ac{RNN} can only give one output at each time step $t$, and to generate grammatical responses, we are actually maximizing token-wise probability at each time step during training:
\begin{equation}
  \textstyle
  \max P(Y|X) = \max \prod_{t=1}^{|Y|} P(y_t| y_{<t}, X),
  \label{eq:obj}
\end{equation}
where $y_{<t} = (y_1, y_2, \dots, y_{t-1})$ are tokens generated in previous time steps.
At test time, the response is generated with respect to:
\begin{equation}
  \textstyle
  \hat{Y} = \argmax_Y P(Y|X).
  \label{eq:obj_test}
\end{equation}
In practice, we usually maximize the aforementioned conditional probability by minimizing the prediction loss at each step $t$, which is actually the \acf{CE} loss:
\begin{equation}
  \label{eq:CE_loss}
  \textstyle  
  \CE(y_t) = -\sum_{i=1}^N \delta_i(y_t) \log(P(c_i|y_{<t}, X)),
\end{equation}
where $(c_1, c_2, \dots, c_N)$ is the search space of $y_t$, $\delta_i(y_t) = 1$ if $y_t = c_i$ and $0$ otherwise; $P(c_i|\cdot)$ is the predicted probability of candidate token $c_i$ and is calculated using the softmax function:
\begin{equation}
  \label{eq:prob}
  P(c_i|y_{<t}, X) =
  \frac{\exp(f_\theta(h_{t-1}^{\mathit{dec}}, y_{t-1}, c_i, X))}{\sum_{j=1}^N
    \exp(f_\theta(h_{t-1}^{\mathit{dec}}, y_{t-1}, c_j, X))},
\end{equation}
where $f_\theta(\cdot)$ is a non-linear scoring function with parameters $\theta$; $f_\theta(\cdot)$ takes the hidden state $h_{t-1}^{\mathit{dec}}$, last generation $y_{t-1}$ and $X$ as inputs, and calculates a score for each possible candidate $c_i$.

\begin{table}[t!]
  \centering
  \caption{Frequency ranks of example tokens and their percentage (\%) in validation and test model outputs, and ground truth outputs of the training data.}
  \label{tab:lm}
  \begin{tabular}{lccccccccc}
    \toprule
    & . & i & 'm & not & n't \\
    \midrule
    Validation rank & 1 (11) & 2 (10) & \phantom{1}3 (5) & \phantom{1}4 (5) & \phantom{1}8 (3) \\
    Test rank & 1 (10) & 2\phantom{1} (9) & \phantom{1}3 (5) & \phantom{1}4 (5) & \phantom{1}8 (3) \\
    Training rank & 1 \phantom{1}(7) & 4\phantom{1} (3) & 35 (1) & 32 (1) & 13 (1) \\
    \bottomrule
  \end{tabular}
\end{table}

During training, the loss calculated using Eq.~\eqref{eq:CE_loss} is back-propa\-gated through the whole network.
The effect is that $f_\theta$ will assign higher scores for $c_i = y_t$ in the future, so that the loss in Eq.~\eqref{eq:CE_loss} will decrease, and meanwhile the predicted probability in Eq.~\eqref{eq:prob} will increase.
The ultimate effect is to maximize the probabilities of ground-truth outputs given input sequences, as illustrated in Eq.~\eqref{eq:obj}.
The model is frequently penalized by a small number of frequent tokens, as a result of which the \ac{TL} for these tokens is higher than for less-frequent ones:
\begin{equation}
  \textstyle
  \TL(c_i) = \sum_{t=1}^{N_t} \CE(y_t = c_i),
  \label{eq:total_loss}
\end{equation}
where $t$ denotes the time step, with maximum training steps $N_t$:
\begin{equation}
  \textstyle
  \label{eq:total_freq}
  N_t = \sum_{i=1}^N \freq(c_i).
\end{equation}
Here, $\freq(c_i)$ represents the frequency that $c_i$ appears in the training ground truth.
Assume that the expected value $\mathbb{E}[\CE(c_i)]$ is roughly the same for $c_i, \forall i \in \{1, 2, \dots, N\}$, then tokens with a higher frequency will have a larger total loss during training.
We refer to this phenomenon as \emph{loss imbalance}.

Due to loss imbalance, a \ac{Seq2Seq} model favors frequent tokens and thus is over-confident about them.
This is especially true for the leading token as observed in Table~\ref{tab:freq}, since the decoder language model does not have a strong effect in the beginning of prediction.
When a frequent token is selected as leading token, the search space for subsequent tokens is hugely restricted by the language model, and this will likely result in a frequent \emph{generic} response.



\section{Frequency-Aware Cross-Entropy Loss}
Now that we have identified \emph{loss imbalance} as a cause of low-diversity problems, we propose to balance the total loss for each token by applying a weight factor to \ac{TL}:
\begin{equation}
  \label{eq:weighted_TL}
  \textstyle
  \WTL(c_i) = w_i \sum_{t=1}^{N_t} CE(y_t = c_i),
\end{equation}
where $w_i$ is the weight corresponding to $c_i$.
By absorbing $w_i$ into the \ac{CE} loss function, we obtain the \ac{FACE} loss function:
\begin{equation}
  \label{eq:wlf}
  \textstyle  
  \FACE(y_t) = -\sum_{i=1}^N w_i \delta_i(y_t) \log(P(c_i|y_{<t}, X)).
\end{equation}
Our key solution to the low-diversity problem lies in the weight factors $w_i$.
Based on the analysis in \S\ref{analysis}, one straightforward way to learning $w_i$ is to take advantage of the token frequency $\freq(c_i)$, so that frequent tokens will have lower weights.
Below we propose two methods to estimate $\freq(c_i)$: \emph{ground-truth or GT frequency} and \emph{output frequency}. 

\subsubsection*{GT frequency: Token frequency in the ground-truth respons\-es.}
As illustrated in Eq.~\eqref{eq:total_freq}, the number of training steps $N_t$ equals the sum of frequencies of tokens in the training ground-truth responses, so it is intuitive to use these token frequencies to adjust the weight in Eq.~\eqref{eq:wlf}.
However, during training, the model is given data sequentially in random order.
As a result, the real-time token frequency seen by the model is likely to differ from that of the entire training data.
Our solution is to calculate the \emph{batch} token frequency instead:
\begin{equation}
  \label{eq:freq}
  \textstyle
  \freq_b(c_i) = \freq_{b-1}(c_i) + \freq_o(c_i),
\end{equation}
where $b$ is the number of training batches seen so far and $o$ represents the newly \emph{observed} batch.

\subsubsection*{Output frequency: Token frequency in model responses.}
Alternatively, we can employ a train-and-refine strategy: the responses of a pre-trained Seq2Seq model can reflect which tokens the model is already overfitted for; by directly penalizing those tokens with a fine-tuning procedure, we can improve the response diversity without retraining it from scratch.
The output frequency may have a more obvious effect on improving response diversity than the GT frequency, because diversity is directly exhibited by model outputs.

\subsection{Weight calculation}
Given the frequency of each token, we introduce the following two methods to calculate the weight factor.

\subsubsection{Pre-weight.} 
This method derives the weight factor $w_i$ prior to seeing new training examples, i.e., \emph{pre-weight function}:
\begin{equation}
  \label{eq:pre}
  w_i = a \times RF_i + 1.
\end{equation}
Here we formulate it as a linear function of \emph{relative frequency}: $RF_i = \freq(c_i)/\sum_j \freq(c_j)$, and $a = -1 / \max RF_j , \forall j \in \{1, \dots, N\}$, is the slope, and the bias is $1$ so that $w_i$ falls in $[0, 1]$.
We then normalize $\{w_1, w_2, \dots, w_N\}$ to have a mean of $1$.
The pre-weight function can make sure that tokens with a higher $RF$ value will get lower weights.
In other words, the influence of high frequency tokens will be penalized by using the pre-weight function.

\subsubsection{Post-weight.} 
This method tries to penalize the model's \emph{conservativeness}: if the output token $y_t$ has a higher frequency than the ground truth $c_i$, which indicates the model conservatively picked a ``safe'' token, then its loss will be scaled up by $w_i > 1$, otherwise $w_i = 1$ due to ReLU activations:
\begin{equation}
  \label{eq:post}
  w_i = 1 + \frac{ReLU(\freq(y_t) - \freq(c_i))}{\sum_j^N \freq(c_j)}.
\end{equation}
Since we can only apply this weighting function after obtaining the model outputs, we refer to it as the \emph{post-weight function}.

\subsection{Empowered by confidence penalty}
As illustrated in \citep{jiang2018why}, \acfi{CP}\acused{CP} methods can alleviate the low-diversity problem.
In the experiments of this paper, we also test the performance of the \ac{CP} function \citep{jiang2018why}:
\begin{equation}
  \label{eq:cp}
  \CP(y_t) = \CE(y_t) - \beta H(p(y_t|y_{<t}, X)),
\end{equation}
where $p(y_t|y_{<t}, X)$ is the predicted distribution at $t$, and $H(\cdot)$ is its entropy.
However, during experiments we found that the parameter $\beta$ needs to be carefully chosen, otherwise the loss $\CP(y_t)$ will be negative, which is counter-intuitive since losses should be greater than 0.
Instead, we propose a \emph{parameter-free} \ac{CP} function:
\begin{equation}
  \label{eq:cpf}
  \CP\nolimits_{\mathit{free}}(y_t) = \CE(y_t) + \frac{1}{H(p(y_t|y_{<t}, X))}.
\end{equation}

\noindent%
\ac{FACE} and \ac{CP} can be easily combined to further improve response diversity.
A trivial combination is to replace the \ac{CE} loss function with the $\FACE$ loss function in Eq.~\eqref{eq:cp} and Eq.~\eqref{eq:cpf}.
We also propose a \ac{CP} weighting function using the entropy in Eq.~\eqref{eq:cpf}:
\begin{equation}
  \label{eq:wlf_cp}
  w = 1 + \frac{1}{H(p(y_t|y_{<t}, X))},
\end{equation}
where $w$ (without subscript) is assumed to be independent of $c_i$, which is different from that in Eq.~\eqref{eq:pre} and Eq.~\eqref{eq:post}.
By using $w$ as the weight of \ac{FACE} in Eq.~\eqref{eq:wlf}, we can penalize the model confidence by adjusting the weight of \ac{FACE}.



\section{Experimental Setup}

To prove the effectiveness of our proposed methods, we design experiments to answer the following questions:
\begin{inparaenum}[(Q1)]
\item Which combination of our proposed frequency methods (GT frequency and output frequency) and weighting functions (pre- and post-weight) performs best?
\item Does \ac{FACE} improve the diversity of \ac{Seq2Seq} conversation models?
\item Does \ac{CP} improve the diversity of \ac{Seq2Seq} conversation models?
\item Does the combination of \ac{FACE} and \ac{CP} further improve performance?
\item Does \ac{FACE} improve the response quality besides diversity?
\end{inparaenum}

\subsection{Baselines and datasets}

\begin{table}[t!]
  \caption{Components of different models.}
  \label{tab:comp}
  \centering
  \begin{tabular}{lcccc}
    \toprule
    Model & Greedy & BS & \#Attn & Extras \\
    \midrule
    Seq2Seq & Yes & No & 1 & No \\
    MMI & No & Yes & 1 & Reverse\footnotemark \\
    MHAM & Yes & No & 5 & No \\
    \midrule
    FACE & Yes & No & 1 & No \\
    CP & Yes & No & 1 & No \\
    \bottomrule
  \end{tabular}
\end{table}
\footnotetext{The MMI-bidi method needs a reverse model which is pre-trained using inverse training examples: (response, message).}

The first baseline is a vanilla Seq2Seq model with \emph{general} attention \citep{luong2015effective} as implemented within the ParlAI platform \citep{miller2017parlai}.
We also choose the \ac{MMI} models proposed in \citep{li2016diversity} and the \ac{MHAM} models proposed in \citep{tao2018get}.
These models are all deterministic methods.
In Table~\ref{tab:comp}, we list the main differences between our methods and the baselines in terms of greedy decoder, beam search (BS), number of attention layers (\#Attn), and other mandatory components (Extras).
To allow for a fair comparison, we implement all our methods and baselines using ParlAI.
We choose two publicly available benchmark datasets: \ac{OSDb} and Twitter, for evaluating the baselines and our proposed methods.
We follow \citet{li2016diversity,tao2018get} to use short-history conversations; see below for details.

\subsubsection{\ac{OSDb}.} 
The \ac{OSDb} dataset \citep{tiedemann2009news} is an online-available corpus of movie subtitles.\footnote{http://www.opensubtitles.org/}
Here, we use the 2011 version and set aside $\sim$60M lines, which constitutes $\sim$30M message-response pairs for training.
Following \citet{li2016diversity}, we randomly select 2K pairs from the IMSDB dataset \citep{walker2011perceived} for validation and test sets, respectively, and we filter out pairs whose responses or messages are shorter than 6 tokens.

\subsubsection{Twitter.} 
Twitter is a commonly used source of data in dialogue generation research.
For ease of reproducibility of the results reported in this paper, we did not follow \citet{sordoni2015neural,li2016diversity} who used $\sim$130M context-message-response triples before pre-processing.
Instead, we use the version released by \citep{serban2017hierarchical} for training, which contains $\sim$4M tweet IDs in total and can be scraped in 3 days.\footnote{On July 11 2018, when we finished scraping, only 2.6M IDs in total are still valid.}
After formatting the tweets as \emph{context-message-response} triples, we have 904K training examples and we concatenate context-message as input.
For the training, validation and test sets, we exclude all overlapping IDs.
Then we restrict the sequence length in both validation and test sets to the range [6, 18], resulting in 18,162 validation and 1,897 test triples.
Furthermore, since tuning the \ac{MMI} models is quite time-consuming, we randomly select 2K triples from the validation set for hyper-parameter selection.

\subsection{Evaluation}
\subsubsection{Automatic evaluation.}
Although reported not to correlate well with human judgments \citep{liu2016not}, we still report BLEU scores \citep{papineni2002bleu} for fair comparisons with baselines.
To evaluate response diversity, we use the \emph{d-1} and \emph{d-2} metrics proposed in \citep{li2016diversity} that are calculated as the number of distinct uni- and bigrams, divided by the total number of tokens generated.

\subsubsection{Human evaluation.}
Following \citet{sordoni2015neural}, we recruit crowd-source workers to perform pairwise qualitative comparisons.
Since the conversation history of the \ac{OSDb} dataset is only a single turn, which is hard for human annotators to judge, we choose the Twitter dataset for human evaluation, which has two-turn histories.
We randomly select 1K test examples from the Twitter test set, and paired model responses (\emph{FACE} vs.\ \emph{baseline}) are shown to 3 evaluators in random order.
Evaluators are told to choose a better response in terms of \emph{relevance}, \emph{interestingness} and \emph{grammar}.
Ties are allowed.
We then carry out Welch's t-test on the human preferences obtained in this manner.

\subsection{Implementation details}\label{params}
For the \ac{OSDb} and Twitter datasets, we keep the most frequent 25,000 tokens and replace other tokens with \emph{\_UNK\_}.
For the \ac{OSDb} corpus, we use a 4-layer \ac{LSTM} network for both encoder and decoder.
The \ac{HS} and word embedding size are set to 1,000.
On the smaller Twitter corpus, we use a smaller network with a 2-layer \acp{LSTM} for encoder and decoder.
The \ac{HS} and word embedding sizes are set to 512 and 200, respectively.
For both networks, we randomly initialize the model parameters from a uniform distribution $\mathcal{U}(-\sqrt{1/\ac{HS}},\sqrt{1/\ac{HS}})$ and the word embeddings from a normal distribution $\mathcal{N}(0, 1)$.
We also employ dropout \citep{srivastava2014dropout} with drop ratio $p$ = 0.1.
We use Adam \citep{kingma2014adam} as our optimization method. 
For the hyper-parameters of the Adam optimizer, we set the learning rate $\alpha = 0.001$, two momentum parameters $\beta1 = 0.9$ and $\beta2 = 0.999$, respectively, and $\epsilon$ = $10^{-8}$. 
We also clip gradients \citep{pascanu2013difficulty} to $5$ during training to avoid gradient explosion.
To speed up training and convergence, we use mini-batches of size 256.
Since the model chosen with minimum training loss usually has very low diversity, we choose \emph{d-1} as the early stopping criterion.
A scheduler is used to reduce the learning rate by a factor of $0.5$ when a \emph{d-1} plateau is detected with $patience=3$.
$\beta$ in Eq.~\eqref{eq:cp} is set to 0.01.



\section{Results and Analyses}

\subsection{\ac{FACE} variants}
To answer Q1, we first identify the best performing variant of~\ac{FACE} to be used in later experiments.
In Tables~\ref{tab:osdb_inner} and~\ref{tab:twitter_inner}, we list the scores for four variants of \ac{FACE}, on the \ac{OSDb} and Twitter datasets, respectively: Output token frequency \& PRe-weight (\emph{FACE-OPR}), Output token frequency \& POst-weight (\emph{FACE-OPO}), GT frequency \& PRe-weight (\emph{FACE-GPR}) and GT frequency \& POst-weight (\emph{FACE-GPO}).

\begin{table}
  \caption{Performance (\%) on the \ac{OSDb} dataset of different variants of \ac{FACE}. Highest scores in \textbf{bold face}.}
  \label{tab:osdb_inner}
  \centering
  \begin{tabular}{lccc}
    \toprule
    Model & d-1 & d-2 & BLEU \\
    \midrule
    \emph{FACE-OPR} & 4.32 & \textbf{20.47} & \textbf{8.03} \\
    \emph{FACE-OPO} & 4.56 & 14.96 & 6.76 \\
    \emph{FACE-GPR} & 2.87 & 10.51 & 7.56 \\
    \emph{FACE-GPO} & \textbf{5.03} & 19.66 & 6.92 \\
    \bottomrule
  \end{tabular}
\end{table}

\begin{table}
  \caption{Performance (\%) on the Twitter dataset of different variants of \ac{FACE}. Highest scores are in \textbf{bold face}.}
  \label{tab:twitter_inner}
  \centering
  \begin{tabular}{lccc}
    \toprule
    Model & d-1 & d-2 & BLEU \\
    \midrule
    \emph{FACE-OPR} & \textbf{6.23} & \textbf{24.18} & 8.33 \\
    \emph{FACE-OPO} & 5.73 & 17.95 & \textbf{8.80} \\
    \emph{FACE-GPR} & 4.13 & 15.03 & 7.99 \\
    \emph{FACE-GPO} & 5.69 & 17.78 & 8.72 \\

    \bottomrule
  \end{tabular}
\end{table}

On the \ac{OSDb} dataset, we can see from Table~\ref{tab:osdb_inner} that \emph{FACE-OPR} and \emph{FACE-GPO} perform better than the other two in terms of diversity scores.
Similar results are shown on the Twitter dataset in Table~\ref{tab:twitter_inner}.
By comparing the results in Table \ref{tab:osdb_inner} and \ref{tab:twitter_inner}, we find that the performance of \emph{FACE-OPR} is very stable.
We therefore choose \emph{FACE-OPR} as our primary model and refer to it as \emph{FACE} in the following sections.
In contrast, \emph{FACE-GPR} performs worst on both datasets, which indicates that GT frequency does not work well with pre-weight.
The BLEU scores in  Table \ref{tab:osdb_inner} and \ref{tab:twitter_inner} show no obvious patterns.

It is worth noting that all variants reported in Table \ref{tab:osdb_inner} and \ref{tab:twitter_inner} are using the train-and-refine strategy: first training the \ac{Seq2Seq} model with \ac{CE}, then fine-tuning it with \ac{FACE}.
We also tried training from scratch using \ac{FACE}.
However, the performance did not show consistent improvements, which is probably because our \emph{equal expected loss} assumption (under Eq. \eqref{eq:total_freq}) is violated in the early training stages.

\subsection{Automatic evaluation}
We turn to Q2, Q3, and Q4.
The automatic evaluation results are shown in Tables~\ref{tab:osdb} and~\ref{tab:twitter}.
\emph{Seq2Seq} is vanilla-Seq2Seq with \emph{general} attention mechanism \citep{luong2015effective}. \emph{Seq2Seq-refine} is a fine-tuned version of \emph{Seq2Seq} with a smaller batch size of 30.
\emph{MMI-antiLM} adjusts the \emph{Seq2Seq} prediction probability using the decoder language model, while \emph{MMI-bidi} utilizes a reverse model.
Both methods use beam size 200.
\emph{MHAM} projects encoder hidden-states to 5 different semantic spaces, and \emph{CMHAM} forces those 5 spaces to be perpendicular to each other.
All of our models are fine-tuned \emph{Seq2Seq} models with a smaller batch size of 30.
\emph{FACE} is fine-tuned using the \ac{FACE} function and output token frequency.
\emph{CP} and \emph{\CPF} are fine-tuned using Eq.~\eqref{eq:cp} and~\eqref{eq:cpf}, respectively.
\emph{FACE-CP} is a linear combination of \ac{FACE} and \ac{CP}.
\emph{FACE-\CPF} is the multiplicative combination of \ac{FACE} and the \ac{CP} weighting function in Eq.~\eqref{eq:wlf_cp}.

\begin{table}
  \caption{Results (\%) on the \ac{OSDb} dataset.
    Highest scores in each column are highlighted using \textbf{bold face}.}
  \label{tab:osdb}
  \centering
  \begin{tabular}{lccc}
    \toprule
    Model & d-1 & d-2 & BLEU \\
    \midrule
    \emph{Seq2Seq} & 2.70 & \phantom{2}8.63 & 7.34 \\
    \emph{Seq2Seq-refine} & 3.61 & 14.16 & 7.61 \\
    \midrule
    \emph{MMI-antiLM} & 2.73 & 10.68 & 7.83 \\
    \emph{MMI-bidi} & 3.06 & 12.19 & 7.01 \\
    \emph{MHAM} & 3.03 & \phantom{2}9.47 & 7.13 \\
    \emph{CMHAM} & 4.10 & 12.92 & 6.88 \\
    \midrule
    \emph{FACE} & 4.32 & \textbf{20.47} & \textbf{8.03} \\
    \emph{CP} & 4.18 & 15.59 & 7.27 \\
    \emph{\CPF} & \textbf{5.48} & 18.59 & 7.08 \\
    \emph{FACE-CP} & 4.63 & 18.32 & 7.89 \\
    \emph{FACE-\CPF} & 4.69 & 18.67 & 6.98 \\
    \bottomrule
  \end{tabular}
\end{table}

As shown in Table \ref{tab:osdb}, all the methods proposed in this paper outperform the baselines in terms of diversity metrics on the \ac{OSDb} dataset.
\emph{FACE} achieves the highest d-2 score (increase of 6.3\%).
\emph{FACE-CP} and \emph{FACE-\CPF} achieve slightly higher d-1 and lower d-2 scores than \emph{FACE}, which can be viewed as the trade-off between \emph{FACE} and \emph{CP}'s.
Although the penalty strength $\beta$ of \emph{CP} is carefully selected, it is interesting to see that our hyper-parameter free method \emph{\CPF} outperforms \emph{CP} by a large margin and achieves the highest d-1 score (1.9\% improvement), which demonstrates the effectiveness of Eq.~\eqref{eq:cpf}.
\emph{FACE} performs best in terms of BLEU scores, which suggests that \emph{FACE} can generate higher-quality responses.

Although the \ac{MMI}-based and \ac{MHAM}-based methods all exhibit various improvements over \emph{Seq2Seq} in terms of d-1 and d-2 scores, they mostly perform worse than \emph{Seq2Seq-refine}.
This suggests that on the \ac{OSDb} dataset, carefully designed \ac{MMI}-based and \ac{MHAM}-based methods are not able to outperform a fine-tuned \emph{Seq2Seq} baseline.

\begin{table}
  \caption{Results (\%) on the Twitter dataset. Highest scores in each
  column are highlighted using \textbf{bold face}.}
  \label{tab:twitter}
  \centering
  \begin{tabular}{lccc}
    \toprule
    Model & d-1 & d-2 & BLEU \\
    \midrule
    \emph{Seq2Seq} & 5.87 & 17.73 & 8.77 \\
    \emph{Seq2Seq-refine} & 5.69 & 17.54 & 8.82 \\
    \midrule
    \emph{MMI-antiLM} & \textbf{6.23} & 18.21 & 6.51 \\
    \emph{MMI-bidi} & 5.42 & 15.16 & 8.20 \\
    \emph{MHAM} & 5.52 & 17.04 & \textbf{8.96} \\
    \emph{CMHAM} & 4.99 & 14.91 & 8.71 \\
    \midrule
    \emph{FACE} & \textbf{6.23} & \textbf{24.18} & 8.33 \\
    \emph{CP} & 5.97 & 18.15 & 8.84 \\
    \emph{\CPF} & 6.00 & 18.67 & 8.82 \\
    \emph{FACE-CP} & 6.07 & 23.50 & 8.25 \\
    \emph{FACE-\CPF} & 5.89 & 17.81 & 8.85 \\
    \bottomrule
  \end{tabular}
\end{table}

\begin{figure}
  \begin{subfigure}{.35\linewidth}
    \centering
    \includegraphics[width=.95\linewidth]{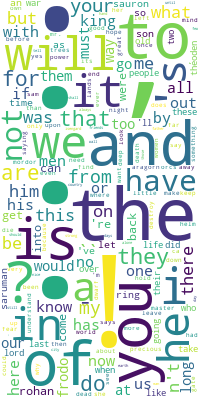}
  \end{subfigure}
  \begin{subfigure}{.35\linewidth}
    \centering
    \includegraphics[width=.95\linewidth]{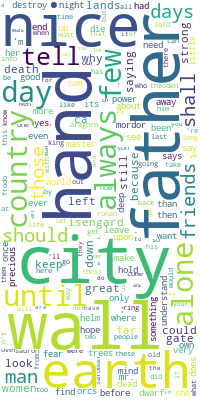}
  \end{subfigure}
  \caption{Word cloud showing top-200 frequent tokens of model responses on the \ac{OSDb} dataset. Left: the larger the font, the higher the frequency. Right: the larger the font, the higher the weight (pre-weight used). (Best viewed in color.)}
  \label{fig:freq_weight}
\end{figure}

Similar to Table \ref{tab:osdb}, Table \ref{tab:twitter} shows that all our methods can increase the diversity of \emph{Seq2Seq} on the Twitter dataset. 
\emph{FACE} achieves the highest d-1 score (0.4\% increase) together with \emph{MMI-antiLM}, but the highest d-2 score of \emph{FACE} (6.5\% improvement) demonstrates that our method fares better.

In Table~\ref{tab:twitter} however, \emph{Seq2Seq-refine} decreases the diversity of \emph{Seq2Seq}, indicating that fine-tuning does not help to address the low-diversity problem on the Twitter dataset.
This is probably because the diversity of \emph{Seq2Seq} is already relatively high on this dataset, and there is limited space for further improvements.

While \emph{MMI-bidi} performs better than \emph{MMI-antiLM} on the \ac{OSDb} dataset, on the Twitter dataset, however, \emph{MMI-bidi} performs much worse and degrades the diversity of the \emph{Seq2Seq} model.
The reason for this phenomenon is probably because of the reverse model of \emph{MMI-bidi}: the input consisting of Twitter triples contains two turns from different speakers (\emph{context-message}), thus given the \emph{response} only, the prediction probability of \emph{context-message} is very unreliable.

Although it achieves the highest BLEU score, \emph{MHAM} slightly hurts diversity.
Similarly, \emph{CMHAM} degrades the diversity even more.
Closer inspection reveals that most of the attention weights of both methods are on the first several tokens of the input sequence (i.e., \emph{context}).
It is unlikely to be able to properly learn the relation between \emph{context-message} and \emph{response} by only paying attention to the \emph{context}.

To illustrate the effectiveness of the weighting function, we display the frequencies and the corresponding weights of some tokens in Figure~\ref{fig:freq_weight}.
From this figure, we can see that the most frequent tokens have very small weights, and less frequent tokens receive larger weights.
Please note that the tokens with the largest weights are in the scope of top-200 frequent tokens.

\subsection{Human evaluation}

\begin{table}
  \caption{Results of the pairwise human evaluation (\%) on the Twitter dataset. ``Win'', ``Lose'' and ``Gain'' correspond to ``FACE wins'', ``Baseline wins'' and their difference ($Win - Lose$), respectively. Highest scores are highlighted in \textbf{bold face}, and $^*, ^{**}, ^{***}$ symbols indicate significant improvements with $p\mbox{-}value < 0.05, <0.01, <0.005$, respectively.}
  \label{tab:human_eval}
  \centering
  \begin{tabular}{llll}
    \toprule
    Comparison & Win & Lose & Gain \\
    \midrule
    \emph{FACE vs Seq2Seq} & \textbf{38.61}$^{***}$ & 21.54 & 17.07 \\
    \emph{FACE vs MMI-antiLM} & \textbf{51.30}$^{***}$ & 19.35 & 31.95\\
    \emph{FACE vs MMI-bidi} & \textbf{61.91}$^{***}$ & 20.92 & 40.99 \\
    \emph{FACE vs MHAM} & \textbf{50.93}$^{**}$ & 42.56 & \phantom{1}8.37 \\
    \emph{FACE vs CMHAM} & \textbf{43.75}$^{*}$ & 38.85 & \phantom{1}4.90 \\
    \bottomrule
  \end{tabular}
\end{table}

We now turn to Q5.
Human evaluation results are reported in Table \ref{tab:human_eval}.
We can see that \emph{FACE} is significantly better than all baselines.
This means that by increasing response diversity, \ac{FACE} can improve relevance and interestingness of responses, without sacrificing grammatical accuracy.
Specifically, by penalizing frequent tokens, \emph{FACE} gives more opportunities to less frequent tokens, so the interestingness is higher than \emph{Seq2Seq}.
While high-frequency tokens may be appropriate responses in more situations than low-frequency ones, it is for the same reason that higher frequency tokens convey less information, thus making the responses seem generic.
Therefore, by encouraging low-frequency tokens, \emph{FACE} has the potential to increase relevance of responses.
By looking at the human evaluation scores in Table \ref{tab:human_eval} together with the BLEU scores in Table \ref{tab:twitter}, we can re-confirm that BLEU scores do not correlate well with human qualitative evaluation \cite{liu2016not}.

Since \emph{FACE} learns a language model relying entirely on training data, the grammatical accuracy of \emph{FACE} is as least as good as that of \emph{Seq2Seq}.
The examples in Table \ref{tab:eg} show that \emph{FACE} can actually generate more relevant and interesting responses.
In contrast, although \emph{MMI-antiLM} improves response diversity, it increases the risk of grammatical errors since it penalizes the language model.
For example, the response of \emph{MMI-antiLM} in the second example of Table \ref{tab:eg} is ungrammatical and irrelevant to the conversation history.
Similarly, \emph{MMI-bidi} uses a reverse model to adjust the model outputs, in the hope that the responses can be more diverse and relevant.
However, as explained earlier, the reverse model is highly unreliable, making many of the responses look weird, see, e.g., the third row of Table \ref{tab:eg}.

Although \emph{MHAM} and \emph{CMHAM} can improve response diversity to some extent, it is not guaranteed that the multiple attention heads are well distributed throughout the whole history.
In fact, there are many cases where the attention weights are on the \textbf{Context} part of the history, making the responses irrelevant to the \textbf{Message}, such as the fourth and fifth examples in Table \ref{tab:eg}.

\begin{table}
  \caption{Example responses from different models.}
  \footnotesize
  \label{tab:eg}
  \centering
  \begin{tabular}{@{}p{0.5\linewidth}p{0.45\linewidth}@{}}
    \toprule
    History & Responses \\
    \midrule
    \textbf{Context}: i need 4 mo followers .. & \emph{\textbf{FACE}}: no i do n't think so .. i 'm watching it now .. \\
    \textbf{Message}: u ever watched criminal minds suspect behavior?? & \emph{\textbf{Seq2Seq}}: i do n't know what to say ..\\
    \midrule
    \textbf{Context}: dont you dare steal my idea. i call shenanigans & \emph{\textbf{FACE}}: well then you should have said it like that . \\
    \textbf{Message}: Totally not stealing. Everyone collaborated on that idea. & \emph{\textbf{MMI-antiLM}}: come on . come on go back to you* \\
    \midrule
    \textbf{Context}: I never watched Spongebob's movie tho, was it good? You know Hey Arnold? & \emph{\textbf{FACE}}: haha , i 'm watching it now , it 's so good : d \\
    \textbf{Message}: yes it is good u should watch :D haha hey arnould with footballegg face :P & \emph{\textbf{MMI-bidi}}: oh , that \\
    \midrule
    \textbf{Context}: perhaps a recipe is in order? I have room in my fridge... ;) & \emph{\textbf{FACE}}: sounds good ! i 'll be sure to check it out ! thanks for sharing the mocha cakes with the butter cakes ; ) \\
    \textbf{Message}: trying this one for 1st time LOL http://bit.ly/nrFMIU start from bottom recipe \& work up & \emph{\textbf{MHAM}}: i 'm sure you 'll find a recipe for me ! \\
    \midrule
    \textbf{Context}: Don't laugh at my picture! That's not nice. & \emph{\textbf{FACE}}: i 'm not a fan of the picture but it 's not that bad , it 's just a pic of me saying `` i 'm a girl '' haha : - ) \\
    \textbf{Message}: i'm laughing because you actually posted the pic. I believed you:-) lol & \emph{\textbf{CMHAM}}: i 'm not laughing at you . \\
    \bottomrule
  \end{tabular}
\end{table}



\section{Conclusion}
In this work, we have proposed a \acf{FACE} function for tackling the low-diversity problem of Seq2Seq-based conversation models.
Experiments on short-history conversation datasets demonstrate that the \ac{FACE} loss function can effectively improve the diversity and quality of responses.
\ac{FACE} achieves the improvements with minimum modifications to the original \ac{Seq2Seq} model, which makes it flexible to extend.
We also propose a hyper-parameter free \CPF, which exhibits better performance than the original parameter-dependent \ac{CP}.

A limitation of \ac{FACE} is that the learning procedure is not as stable as \ac{CE}, which increases the difficulty of training.
In future work, we would like to investigate this phenomenon in depth.
We also hope to test \ac{FACE} in a long-history conversation setting by applying \ac{FACE} to hierarchical \ac{Seq2Seq} \citep{serban2016building}.
Besides, we also plan to apply \ac{FACE} to stochastic models \citep{serban2017hierarchical,zhao2017learning} and examine how \ac{FACE} can be used in an adversarial dialogue generation setup~\cite{li-dialogue-2019}.


\section*{Acknowledgments}
This research was supported by the China Scholarship Council,
Ahold Delhaize,
the Association of Universities in the Netherlands (VSNU),
and
the Innovation Center for Artificial Intelligence (ICAI).
All content represents the opinion of the authors, which is not necessarily shared or endorsed by their respective employers and/or sponsors.

\bibliographystyle{ACM-Reference-Format}
\balance 
\bibliography{chatbot}


\begin{thebibliography}{19}


\ifx \showCODEN    \undefined \def \showCODEN     #1{\unskip}     \fi
\ifx \showDOI      \undefined \def \showDOI       #1{#1}\fi
\ifx \showISBNx    \undefined \def \showISBNx     #1{\unskip}     \fi
\ifx \showISBNxiii \undefined \def \showISBNxiii  #1{\unskip}     \fi
\ifx \showISSN     \undefined \def \showISSN      #1{\unskip}     \fi
\ifx \showLCCN     \undefined \def \showLCCN      #1{\unskip}     \fi
\ifx \shownote     \undefined \def \shownote      #1{#1}          \fi
\ifx \showarticletitle \undefined \def \showarticletitle #1{#1}   \fi
\ifx \showURL      \undefined \def \showURL       {\relax}        \fi
\providecommand\bibfield[2]{#2}
\providecommand\bibinfo[2]{#2}
\providecommand\natexlab[1]{#1}
\providecommand\showeprint[2][]{arXiv:#2}

\bibitem[\protect\citeauthoryear{Jiang and de~Rijke}{Jiang and
  de~Rijke}{2018}]%
        {jiang2018why}
\bibfield{author}{\bibinfo{person}{Shaojie Jiang} {and}
  \bibinfo{person}{Maarten de Rijke}.} \bibinfo{year}{2018}\natexlab{}.
\newblock \showarticletitle{Why are Sequence-to-Sequence Models So Dull?
  Understanding the Low-Diversity Problem of Chatbots}. In
  \bibinfo{booktitle}{\emph{Proceedings of the 2018 EMNLP Workshop SCAI: The
  2nd International Workshop on Search-Oriented Conversational AI}}
  \emph{(\bibinfo{series}{SCAI '18})}.
\newblock


\bibitem[\protect\citeauthoryear{Kingma and Ba}{Kingma and Ba}{2015}]%
        {kingma2014adam}
\bibfield{author}{\bibinfo{person}{Diederik~P Kingma} {and}
  \bibinfo{person}{Jimmy Ba}.} \bibinfo{year}{2015}\natexlab{}.
\newblock \showarticletitle{Adam: A Method for Stochastic Optimization}. In
  \bibinfo{booktitle}{\emph{International Conference on Learning
  Representations}} \emph{(\bibinfo{series}{ICLR '15})}.
\newblock


\bibitem[\protect\citeauthoryear{Li, Galley, Brockett, Gao, and Dolan}{Li
  et~al\mbox{.}}{2016}]%
        {li2016diversity}
\bibfield{author}{\bibinfo{person}{Jiwei Li}, \bibinfo{person}{Michel Galley},
  \bibinfo{person}{Chris Brockett}, \bibinfo{person}{Jianfeng Gao}, {and}
  \bibinfo{person}{Bill Dolan}.} \bibinfo{year}{2016}\natexlab{}.
\newblock \showarticletitle{A Diversity-Promoting Objective Function for Neural
  Conversation Models}. In \bibinfo{booktitle}{\emph{Proceedings of the 2016
  Conference of the North American Chapter of the Association for Computational
  Linguistics: Human Language Technologies}} \emph{(\bibinfo{series}{NAACL-HLT
  '16})}. \bibinfo{pages}{110--119}.
\newblock


\bibitem[\protect\citeauthoryear{Li, Kiseleva, and de~Rijke}{Li
  et~al\mbox{.}}{2019}]%
        {li-dialogue-2019}
\bibfield{author}{\bibinfo{person}{Ziming Li}, \bibinfo{person}{Julia
  Kiseleva}, {and} \bibinfo{person}{Maarten de Rijke}.}
  \bibinfo{year}{2019}\natexlab{}.
\newblock \showarticletitle{Dialogue generation: From imitation learning to
  inverse reinforcement learning}. In \bibinfo{booktitle}{\emph{AAAI 2019: 33rd
  AAAI Conference on Artificial Intelligence}}. \bibinfo{publisher}{AAAI}.
\newblock


\bibitem[\protect\citeauthoryear{Liu, Lowe, Serban, Noseworthy, Charlin, and
  Pineau}{Liu et~al\mbox{.}}{2016}]%
        {liu2016not}
\bibfield{author}{\bibinfo{person}{Chia{-}Wei Liu}, \bibinfo{person}{Ryan
  Lowe}, \bibinfo{person}{Iulian Serban}, \bibinfo{person}{Michael Noseworthy},
  \bibinfo{person}{Laurent Charlin}, {and} \bibinfo{person}{Joelle Pineau}.}
  \bibinfo{year}{2016}\natexlab{}.
\newblock \showarticletitle{How {NOT} To Evaluate Your Dialogue System: An
  Empirical Study of Unsupervised Evaluation Metrics for Dialogue Response
  Generation}. In \bibinfo{booktitle}{\emph{Proceedings of the 2016 Conference
  on Empirical Methods in Natural Language Processing}}
  \emph{(\bibinfo{series}{EMNLP '16})}. \bibinfo{pages}{2122--2132}.
\newblock


\bibitem[\protect\citeauthoryear{Luong, Pham, and Manning}{Luong
  et~al\mbox{.}}{2015}]%
        {luong2015effective}
\bibfield{author}{\bibinfo{person}{Thang Luong}, \bibinfo{person}{Hieu Pham},
  {and} \bibinfo{person}{Christopher~D. Manning}.}
  \bibinfo{year}{2015}\natexlab{}.
\newblock \showarticletitle{Effective Approaches to Attention-based Neural
  Machine Translation}. In \bibinfo{booktitle}{\emph{Proceedings of the 2015
  Conference on Empirical Methods in Natural Language Processing}}
  \emph{(\bibinfo{series}{EMNLP '15})}. \bibinfo{pages}{1412--1421}.
\newblock


\bibitem[\protect\citeauthoryear{Miller, Feng, Batra, Bordes, Fisch, Lu,
  Parikh, and Weston}{Miller et~al\mbox{.}}{2017}]%
        {miller2017parlai}
\bibfield{author}{\bibinfo{person}{Alexander Miller}, \bibinfo{person}{Will
  Feng}, \bibinfo{person}{Dhruv Batra}, \bibinfo{person}{Antoine Bordes},
  \bibinfo{person}{Adam Fisch}, \bibinfo{person}{Jiasen Lu},
  \bibinfo{person}{Devi Parikh}, {and} \bibinfo{person}{Jason Weston}.}
  \bibinfo{year}{2017}\natexlab{}.
\newblock \showarticletitle{ParlAI: A Dialog Research Software Platform}. In
  \bibinfo{booktitle}{\emph{Proceedings of the 2017 Conference on Empirical
  Methods in Natural Language Processing}} \emph{(\bibinfo{series}{EMNLP
  '17})}. \bibinfo{pages}{79--84}.
\newblock


\bibitem[\protect\citeauthoryear{Papineni, Roukos, Ward, and Zhu}{Papineni
  et~al\mbox{.}}{2002}]%
        {papineni2002bleu}
\bibfield{author}{\bibinfo{person}{Kishore Papineni}, \bibinfo{person}{Salim
  Roukos}, \bibinfo{person}{Todd Ward}, {and} \bibinfo{person}{Wei-Jing Zhu}.}
  \bibinfo{year}{2002}\natexlab{}.
\newblock \showarticletitle{BLEU: A Method for Automatic Evaluation of Machine
  Translation}. In \bibinfo{booktitle}{\emph{Proceedings of the 40th Annual
  Meeting on Association for Computational Linguistics}}
  \emph{(\bibinfo{series}{ACL '02})}. \bibinfo{pages}{311--318}.
\newblock


\bibitem[\protect\citeauthoryear{Pascanu, Mikolov, and Bengio}{Pascanu
  et~al\mbox{.}}{2013}]%
        {pascanu2013difficulty}
\bibfield{author}{\bibinfo{person}{Razvan Pascanu}, \bibinfo{person}{Tomas
  Mikolov}, {and} \bibinfo{person}{Yoshua Bengio}.}
  \bibinfo{year}{2013}\natexlab{}.
\newblock \showarticletitle{On the Difficulty of Training Recurrent Neural
  Networks}. In \bibinfo{booktitle}{\emph{Proceedings of the 30th International
  Conference on Machine Learning}} \emph{(\bibinfo{series}{ICML '13})}.
  \bibinfo{pages}{1310--1318}.
\newblock


\bibitem[\protect\citeauthoryear{Pereyra, Tucker, Chorowski, Kaiser, and
  Hinton}{Pereyra et~al\mbox{.}}{2017}]%
        {pereyra2017regularizing}
\bibfield{author}{\bibinfo{person}{Gabriel Pereyra}, \bibinfo{person}{George
  Tucker}, \bibinfo{person}{Jan Chorowski}, \bibinfo{person}{{\L}ukasz Kaiser},
  {and} \bibinfo{person}{Geoffrey Hinton}.} \bibinfo{year}{2017}\natexlab{}.
\newblock \showarticletitle{Regularizing Neural Networks by Penalizing
  Confident Output Distributions}. In \bibinfo{booktitle}{\emph{International
  Conference on Learning Representations}} \emph{(\bibinfo{series}{ICLR '17})}.
\newblock


\bibitem[\protect\citeauthoryear{Serban, Sordoni, Bengio, Courville, and
  Pineau}{Serban et~al\mbox{.}}{2016}]%
        {serban2016building}
\bibfield{author}{\bibinfo{person}{Iulian~Vlad Serban},
  \bibinfo{person}{Alessandro Sordoni}, \bibinfo{person}{Yoshua Bengio},
  \bibinfo{person}{Aaron~C Courville}, {and} \bibinfo{person}{Joelle Pineau}.}
  \bibinfo{year}{2016}\natexlab{}.
\newblock \showarticletitle{Building End-To-End Dialogue Systems Using
  Generative Hierarchical Neural Network Models}. In
  \bibinfo{booktitle}{\emph{Proceedings of the 30th {AAAI} Conference on
  Artificial Intelligence}} \emph{(\bibinfo{series}{AAAI '16})}.
  \bibinfo{pages}{3776--3784}.
\newblock


\bibitem[\protect\citeauthoryear{Serban, Sordoni, Lowe, Charlin, Pineau,
  Courville, and Bengio}{Serban et~al\mbox{.}}{2017}]%
        {serban2017hierarchical}
\bibfield{author}{\bibinfo{person}{Iulian~Vlad Serban},
  \bibinfo{person}{Alessandro Sordoni}, \bibinfo{person}{Ryan Lowe},
  \bibinfo{person}{Laurent Charlin}, \bibinfo{person}{Joelle Pineau},
  \bibinfo{person}{Aaron~C Courville}, {and} \bibinfo{person}{Yoshua Bengio}.}
  \bibinfo{year}{2017}\natexlab{}.
\newblock \showarticletitle{A Hierarchical Latent Variable Encoder-Decoder
  Model for Generating Dialogues}. In \bibinfo{booktitle}{\emph{Proceedings of
  the 31st {AAAI} Conference on Artificial Intelligence}}
  \emph{(\bibinfo{series}{AAAI '17})}. \bibinfo{pages}{3295--3301}.
\newblock


\bibitem[\protect\citeauthoryear{Sordoni, Galley, Auli, Brockett, Ji, Mitchell,
  Nie, Gao, and Dolan}{Sordoni et~al\mbox{.}}{2015}]%
        {sordoni2015neural}
\bibfield{author}{\bibinfo{person}{Alessandro Sordoni}, \bibinfo{person}{Michel
  Galley}, \bibinfo{person}{Michael Auli}, \bibinfo{person}{Chris Brockett},
  \bibinfo{person}{Yangfeng Ji}, \bibinfo{person}{Margaret Mitchell},
  \bibinfo{person}{Jian-Yun Nie}, \bibinfo{person}{Jianfeng Gao}, {and}
  \bibinfo{person}{Bill Dolan}.} \bibinfo{year}{2015}\natexlab{}.
\newblock \showarticletitle{A Neural Network Approach to Context-Sensitive
  Generation of Conversational Responses}. In
  \bibinfo{booktitle}{\emph{Proceedings of the 2015 Conference of the North
  American Chapter of the Association for Computational Linguistics: Human
  Language Technologies}} \emph{(\bibinfo{series}{NAACL-HLT '15})}.
  \bibinfo{pages}{196--205}.
\newblock


\bibitem[\protect\citeauthoryear{Srivastava, Hinton, Krizhevsky, Sutskever, and
  Salakhutdinov}{Srivastava et~al\mbox{.}}{2014}]%
        {srivastava2014dropout}
\bibfield{author}{\bibinfo{person}{Nitish Srivastava},
  \bibinfo{person}{Geoffrey Hinton}, \bibinfo{person}{Alex Krizhevsky},
  \bibinfo{person}{Ilya Sutskever}, {and} \bibinfo{person}{Ruslan
  Salakhutdinov}.} \bibinfo{year}{2014}\natexlab{}.
\newblock \showarticletitle{Dropout: A Simple Way to Prevent Neural Networks
  from Overfitting}.
\newblock \bibinfo{journal}{\emph{The Journal of Machine Learning Research}}
  \bibinfo{volume}{15}, \bibinfo{number}{1} (\bibinfo{year}{2014}),
  \bibinfo{pages}{1929--1958}.
\newblock


\bibitem[\protect\citeauthoryear{Tao, Gao, Shang, Wu, Zhao, and Yan}{Tao
  et~al\mbox{.}}{2018}]%
        {tao2018get}
\bibfield{author}{\bibinfo{person}{Chongyang Tao}, \bibinfo{person}{Shen Gao},
  \bibinfo{person}{Mingyue Shang}, \bibinfo{person}{Wei Wu},
  \bibinfo{person}{Dongyan Zhao}, {and} \bibinfo{person}{Rui Yan}.}
  \bibinfo{year}{2018}\natexlab{}.
\newblock \showarticletitle{Get The Point of My Utterance! Learning Towards
  Effective Responses with Multi-Head Attention Mechanism}. In
  \bibinfo{booktitle}{\emph{Proceedings of the 27th International Joint
  Conference on Artificial Intelligence}} \emph{(\bibinfo{series}{IJCAI '18})}.
  \bibinfo{pages}{4418--4424}.
\newblock


\bibitem[\protect\citeauthoryear{Tiedemann}{Tiedemann}{2009}]%
        {tiedemann2009news}
\bibfield{author}{\bibinfo{person}{J{\"o}rg Tiedemann}.}
  \bibinfo{year}{2009}\natexlab{}.
\newblock \showarticletitle{News from OPUS-A Collection of Multilingual
  Parallel Corpora with Tools and Interfaces}. In
  \bibinfo{booktitle}{\emph{Recent Advances in Natural Language Processing}}
  \emph{(\bibinfo{series}{RANLP '09})}, Vol.~\bibinfo{volume}{5}.
  \bibinfo{pages}{237--248}.
\newblock


\bibitem[\protect\citeauthoryear{Vinyals and Le}{Vinyals and Le}{2015}]%
        {vinyals2015neural}
\bibfield{author}{\bibinfo{person}{Oriol Vinyals} {and}
  \bibinfo{person}{Quoc~V. Le}.} \bibinfo{year}{2015}\natexlab{}.
\newblock \showarticletitle{A Neural Conversational Model}. In
  \bibinfo{booktitle}{\emph{ICML Deep Learning Workshop}}.
\newblock


\bibitem[\protect\citeauthoryear{Walker, Grant, Sawyer, Lin, Wardrip-Fruin, and
  Buell}{Walker et~al\mbox{.}}{2011}]%
        {walker2011perceived}
\bibfield{author}{\bibinfo{person}{Marilyn~A Walker}, \bibinfo{person}{Ricky
  Grant}, \bibinfo{person}{Jennifer Sawyer}, \bibinfo{person}{Grace~I Lin},
  \bibinfo{person}{Noah Wardrip-Fruin}, {and} \bibinfo{person}{Michael Buell}.}
  \bibinfo{year}{2011}\natexlab{}.
\newblock \showarticletitle{Perceived or Not Perceived: Film Character Models
  for Expressive NLG}. In \bibinfo{booktitle}{\emph{International Conference on
  Interactive Digital Storytelling}} \emph{(\bibinfo{series}{ICIDS '09})}.
  \bibinfo{pages}{109--121}.
\newblock


\bibitem[\protect\citeauthoryear{Zhao, Zhao, and Eskenazi}{Zhao
  et~al\mbox{.}}{2017}]%
        {zhao2017learning}
\bibfield{author}{\bibinfo{person}{Tiancheng Zhao}, \bibinfo{person}{Ran Zhao},
  {and} \bibinfo{person}{Maxine Eskenazi}.} \bibinfo{year}{2017}\natexlab{}.
\newblock \showarticletitle{Learning Discourse-level Diversity for Neural
  Dialog Models using Conditional Variational Autoencoders}. In
  \bibinfo{booktitle}{\emph{Proceedings of the 55th Annual Meeting of the
  Association for Computational Linguistics}} \emph{(\bibinfo{series}{ACL
  '17})}. \bibinfo{pages}{654--664}.
\newblock


\end{thebibliography}

\end{document}